\documentclass[twocolumn]{aastex62}
\submitjournal{ApJ}
\accepted{2018-03-11}
\usepackage{epstopdf}

\newcommand{\xmm}{{\it XMM-Newton}}
\newcommand{\xmms}{{\it XMM}}
\newcommand{\chandra}{{\it Chandra}}

\newcommand{\asca}{{\it ASCA}}

\newcommand{\nhunit}{$\mbox{cm}^{-2}$}
\def\farcs{%
 \mbox{%
  \kern  0.13ex.%
  \kern -0.95ex\raisebox{.9ex}{\scriptsize$\prime\prime$}%
  \kern -0.1ex%
 }%
}%

\newcommand{\dem}{DEM L71}
\newcommand{\blobs}{blobs}
\newcommand{\blob}{blob}
\newcommand{\msun}{M$_{\odot}$}
\newcommand{\vdH}{vdH03}
\newcommand{\hughes}{H03}

\newcommand{\maggi}{M16}
\newcommand{\rakowski}{R03}
\newcommand{\ghavamian}{G03}

\shorttitle{SPI Analysis of SNR DEM L71}
\shortauthors{Frank et al.}

\begin{document}
\title{Smoothed Particle Inference Analysis of SNR DEM L71}

\correspondingauthor{Kari A. Frank}
\email{kari.frank@northwestern.edu}

\author[0000-0003-0570-9951]{Kari A. Frank}
\affil{Center for Interdisciplinary Exploration and Research in Astrophysics, Northwestern University \\
2145 Sheridan Rd \\
Evanston, IL 60208, USA}

\author[0000-0002-4661-7001]{Vikram Dwarkadas}
\affiliation{Department of Astronomy and Astrophysics, University of Chicago \\
5640 S Ellis Ave \\
Chicago, IL 60637, USA}

\author{Aldo Panfichi}
\affiliation{Department of Astronomy and Astrophysics, University of Chicago \\
5640 S Ellis Ave \\
Chicago, IL 60637, USA}

\author{Ryan Matthew Crum}
\affiliation{Department of Astronomy and Astrophysics, Pennsylvania State University \\
525 Davey Laboratory \\
University Park, PA 16802, USA}

\author[0000-0003-0729-1632]{David N. Burrows}
\affiliation{Department of Astronomy and Astrophysics, Pennsylvania State University \\
525 Davey Laboratory \\
University Park, PA 16802, USA}

\begin{abstract}
Supernova remnants (SNRs) are complex, three-dimensional objects; properly accounting for this complexity when modeling the resulting X-ray emission presents quite a challenge and makes it difficult to accurately characterize the properties of the full SNR volume. We apply for the first time a novel analysis method, Smoothed Particle Inference, that can be used to study and characterize the structure, dynamics, morphology, and abundances of the entire remnant with a single analysis. We apply the method to the Type Ia supernova remnant DEM L71. We present histograms and maps showing global properties of the remnant, including temperature, abundances of various elements, abundance ratios, and ionization age. Our analysis confirms the high abundance of Fe within the ejecta of the supernova, which has led to it being typed as a Ia. We demonstrate that the results obtained via this method are consistent with results derived from numerical simulations carried out by us, as well as with previous analyses in the literature. At the same time, we show that despite its regular appearance, the temperature and other parameter maps exhibit highly irregular substructure which is not captured with typical X-ray analysis methods.
\end{abstract}



\section{Introduction}
\label{section:intro}
Supernova remnants (SNRs) are invaluable tools for studying stellar evolution, supernovae, and the interstellar medium. As a SNR expands into the interstellar medium (ISM), the shock probes the ambient medium. Heavy elements produced by nucleosynthesis in the progenitor star, and in the explosion, are carried out by the shock wave.  The outgoing fast shock reaches velocities of thousands of kilometers per second, with the resulting post-shock temperatures exceeding millions of degrees. As a result, the best wavelength range to  probe the bulk of the remnant is X-rays. However, SNRs have complex morphology, in 3 dimensions. They also have correspondingly complex spectral properties, typically with significant variation across the face of the remnant, as well as along the line-of-sight. Relative to this spatial and spectral complexity, X-ray observations of SNRs typically provide very few photons. These factors combine to make accurately measuring the physical conditions throughout the entire SNR volume a very challenging endeavor.

The SPIES (Smoothed Particle Inference Exploration of Supernova remnants) project aims to address this challenge by applying a fundamentally different approach to analyzing X-ray observations of SNRs. Smoothed Particle Inference (SPI) was developed specifically for \xmm\ \citep{Peterson2007} and it simultaneously takes advantage of both the spatial and spectral information available in the data. SPI is a Bayesian modeling process that fits a population of gas blobs (``smoothed particles") such that their superposed emission reproduces the observed spatial and spectral distribution of photons. The results of this completely independent approach can be compared with results from numerical simulations as well as results derived from more traditional analysis techniques. Here we demonstrate the unique capabilities of SPI and compare it to more traditional X-ray analysis methods by applying SPI to the Type Ia supernova remnant \dem. 

\dem\ was first identified as an optical SNR in the Large Magellanic Cloud (LMC) more than 40 years ago by 
\citet{daviesetal76}. It was detected in X-rays
in an {\it Einstein} survey by \citet{longetal81}. Since then, it has
been further observed in X-rays using {\it ASCA}
\citep[][]{hughesetal98}, \chandra\ 
\citep[][hereafter \hughes\ and \rakowski]{hughesetal03, rakowskietal03}, and \xmm\ \citep[][hereafter \vdH\ and \maggi]{vanderheydenetal03,Maggi2016}.

The X-ray morphology is remarkably regular, with a bright outer rim and a faint, diffuse center, usually interpreted as the forward shock and reverse-shocked ejecta, respectively. Most previous analyses have therefore characterized the physical conditions in \dem\ with two sets of spectral parameters: one for the outer rim, and one for the center emission. 
\asca\ observations had indicated that the remnant ejecta were enhanced in Fe \citep{hughesetal98},
which was further confirmed by subsequent high-resolution
observations with \chandra\ (\hughes, \rakowski) and \xmm\ (\vdH). These
observations, along with associated Fe ejecta mass estimates of $\sim$1.4\msun (\hughes, \vdH), suggest that the progenitor of \dem\ was a Type Ia supernova (SN),
which are expected to have high Fe abundances. Optical observations of the Balmer-dominated shock velocities by \citet[][hereafter \ghavamian]{ghavamianetal03} yield an age estimate of $\sim$4400 years. Throughout this work, we assume a distance to the LMC and \dem\ of 50 kpc.

\section{Numerical Hydrodynamic Simulations}
\label{section:simulations}

In order to better interpret the results of the SPI analysis, we have carried out spherically symmetric one and two-dimensional numerical hydrodynamic simulations of the evolution of \dem. Analytical approximations can help to understand the overall evolution of the remnant, but are unable to capture the complex dynamics and kinematics. Numerical simulations are necessary to investigate the detailed morphology, deviations from symmetry, and hydrodynamical instabilities. They can be used  to study the temperature, density, and velocity profiles, and the small-scale  structure within the remnant, and to provide an adequate theoretical framework to analyze our SPI results. Results from the these simulations can be used to determine distinguishing properties of the different components of the SNR, such as the contact discontinuity and the ejecta, which can then be used to identify these components in our SPI results.

The explosion of a star to form a supernova (SN) gives rise to a
forward shock that propagates outwards into the medium with a high
velocity, and a reverse shock that moves back into the ejecta in a
Lagrangian sense. The two are separated by a contact discontinuity
(CD) that separates the shocked ejecta from the shocked circumstellar
medium (CSM). The decelerating contact discontinuity is subject to
Rayleigh-Taylor (R-T) instabilities that can result in some mixing of
shocked ejecta into shocked CSM. The density profile of the ejecta and of the CSM are essential to deciphering the shock structure, and kinematics and dynamics of the evolution \citep[see, for instance,][and references within]{vvd11}.

Observations around the rim of the remnant have detected a wide variation in the expansion velocity of the remnant, ranging from
around 400 to about 1250 km s$^{-1}$ \citep{rakowskietal03,Rakowski2009}. These suggest an asymmetry in the
expansion. Our spherically symmetric simulations are not designed to reproduce such an asymmetry; therefore we have not attempted to match all the
parameters. Rather, our goal was to carry out simulations that 
allow us to adequately understand the histograms and maps generated
from the SPI analysis, and provide a framework to interpret the output from
the SPI analysis. Our simulations were run using the VH-1 code, a 1, 2, and 3-dimensional finite-difference numerical hydrodynamics code, based on the Piecewise Parabolic Method of \citet{cw84}. The initial setup is very similar to that used in \citet{dc98} and \citet{vvd00}. Description of the code is available in these papers.

\hughes\ identify the contact discontinuity in \dem\ as
the edge of an inner region of harder X-ray emission, and find that it has
a mean radius of 4.3 pc, which is about halfway out to the forward
shock. The reverse shock will be at a much lower radius. Since the
ejecta are expanding homologously, their density decreases as
t$^{-3}$, which means that the density has decreased substantially,
given the estimated age of the remnant. The reverse shock expanding into this
low density would have moved back into the ejecta,
suggesting that much of the ejecta has been recently shocked and the
reverse shock has recently reached, or is close to, the center of the explosion.

\citet{dc98} have shown that the ejecta density in a Type Ia SN
explosion is best represented by an exponential profile. We used this
profile to describe the SN ejecta expanding into a constant density
medium with a density $\sim$$0.5$ cm$^{-3}$, as suggested by
\ghavamian. Our simulations used an expanding grid, such
that the grid expands outwards with the flow; this is extremely
useful in cases like this where the remnant size grows by 4 orders of
magnitude over the simulation time. The main constraints 
are the radius of the remnant, which is somewhere between 8.6 -- 10
pc, and the forward shock velocities, which are of order a few hundred
km s$^{-1}$.

We first carried out simulations assuming canonical parameters such as
a standard energy of 10$^{51}$ erg, and a mass of 1.4\msun. However,
at the remnant radius of even 10pc, we find that the velocities are
higher than the highest observed velocities, at an age of about 3600
years. This implies either a much higher density, or a lower explosion
energy. Following the results of \ghavamian, we lowered
the energy to 0.4 $\times 10^{51}$ erg, and reran the
simulations. These provide much better agreement with the available
data.

\begin{figure}[htbp]
\begin{center}
\includegraphics[width=0.5\textwidth]{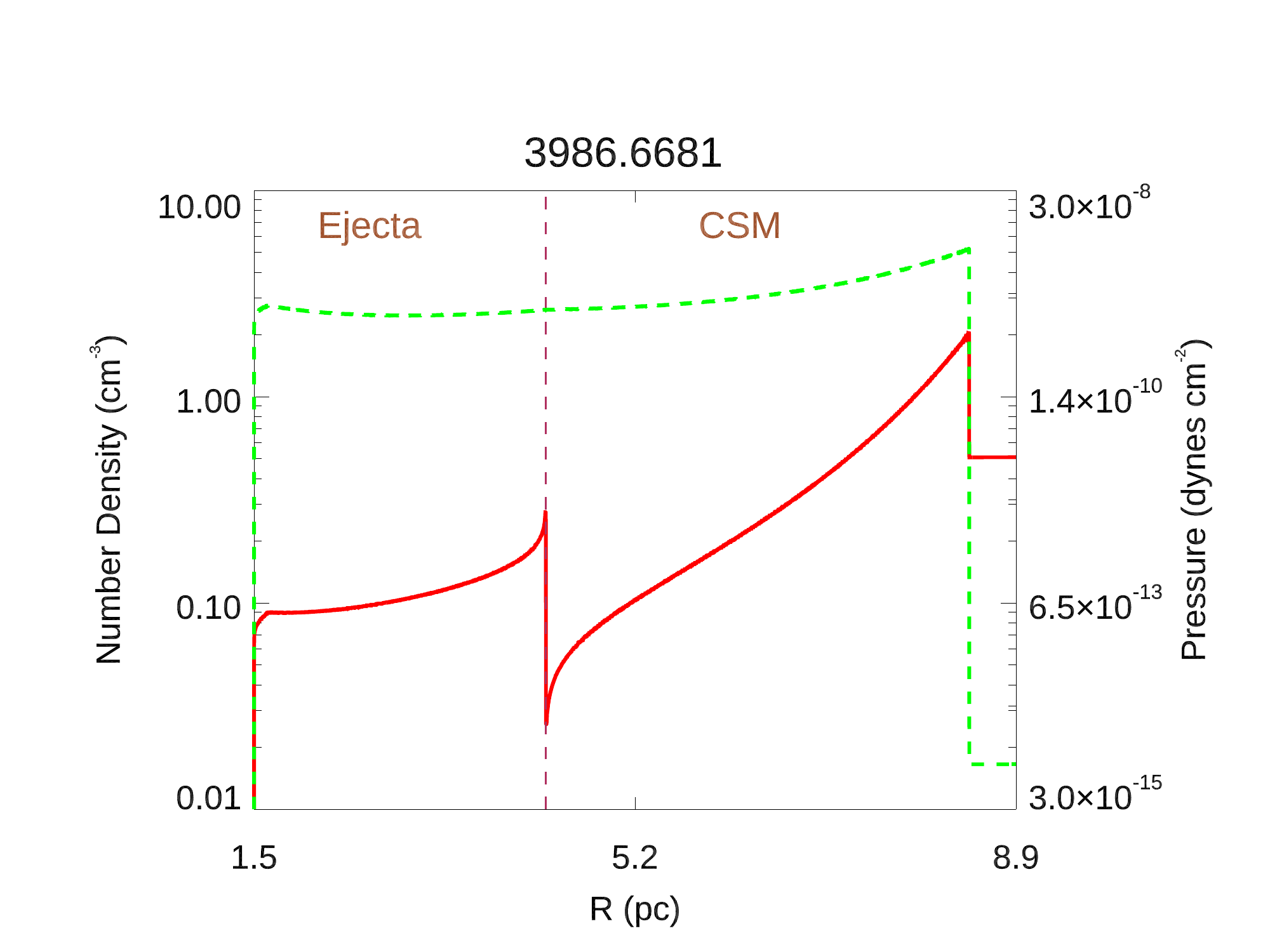}
\caption{\footnotesize The pressure (green) and density (red) profiles
  within the remnant, from the simulation outlined in the text. Time
  in years is given at the top. Density scale is on the left axis, pressure
  scale on the right axis. Number density is calculated by assuming a mean molecular weight of $\approx$ 1.35. }
\label{fig:den}
\end{center}
\end{figure}

In Figure \ref{fig:den} we show the pressure and density profile
within the remnant at an age of 3987 years from our simulation. The
age and the shock velocity are consistent with the results of
\hughes, which is not very surprising since they used a
similar model in their calculations.
 
The forward shock of the remnant is at about 8.5pc. In our expanding grid simulation, the reverse shock
has crossed inward of the inner boundary, and presumably reached the center or is
very close to it. The solution in a few zones
near the inner boundary reflects that and should be ignored. At around
4.3pc lies the contact discontinuity (CD), which separates the shocked
ejecta from the shocked circumstellar medium (CSM). Everything inward of the
CD is ejecta, thus showing that even at this late stage the (shocked)
ejecta are clearly visible. Everything outside the CD is the shocked
CSM. We note that the maximum density is immediately behind
the circumstellar shock. The density decreases inwards till it reaches
the CD, where it increases again, before decreasing towards the
center. The majority of the CSM can be seen to have a
higher density than most of the ejecta. The pressure drops somewhat
behind the outer shock but is then mostly constant throughout the
remnant. The temperature structure will then be essentially the
inverse of the density structure.

\begin{figure}[htbp]
\begin{center}
\includegraphics[width=0.5\textwidth]{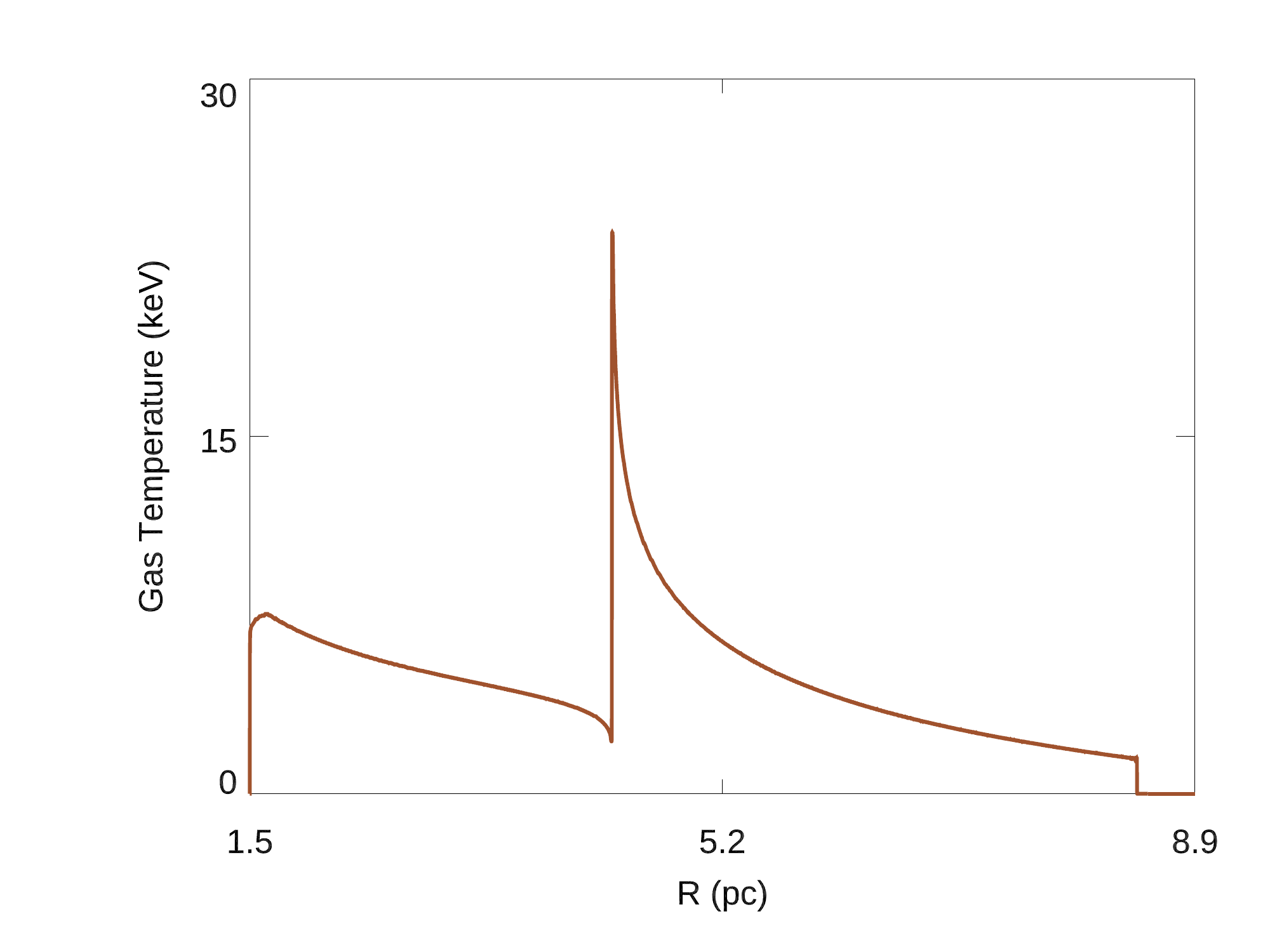}
\caption{\footnotesize The gas temperature within the remnant at an
  age of about 4000 years, from the simulation outlined in the text.}
\label{fig:temp}
\end{center}
\end{figure}

In Figure \ref{fig:temp} we show the gas temperature within the
remnant. It must be emphasized that this is the fluid temperature, and
is obtained by assuming a constant value of the mean molecular weight
throughout the remnant. One would expect that the ejecta would
generally be higher in metals than the CSM and therefore the mean molecular
weight would be different. More importantly, this does not
reflect the electron temperature, which may be lower if electron-ion
equilibration has not been established, as is most likely the case. However, it provides us some
idea of the temperature profile throughout the remnant, and indicates
that on average the ejecta have a higher temperature than the
CSM. There will be a small region of shocked
circumstellar material with a much higher temperature than the ejecta,
but the mass there is very small compared to the total CSM mass.

It is clear that the material nearest to both the shocks forms the most recently
shocked material. This material will consequently have the lowest
ionization age $n_e t$. However, the low ionization age material
behind the outer shock will have essentially the highest density,
whereas the low ionization age material near the reverse shock,
closer to the center of the remnant, will have the lowest
density. Since the ejecta generally have a lower density, they will
have on average lower ionization age than the CSM. Similarly, we
expect that the ejecta, if this is truly a Type Ia remnant as
emphasized by many of the above papers, should display higher
metallicity, and specifically high abundances of Fe and Si as expected
from Type Ia SNe. These distinguishing characteristics can be used to
separate the ejecta from the CSM during the SPI analysis.

\begin{figure}[htbp]
\begin{center}
\includegraphics[width=0.5\textwidth]{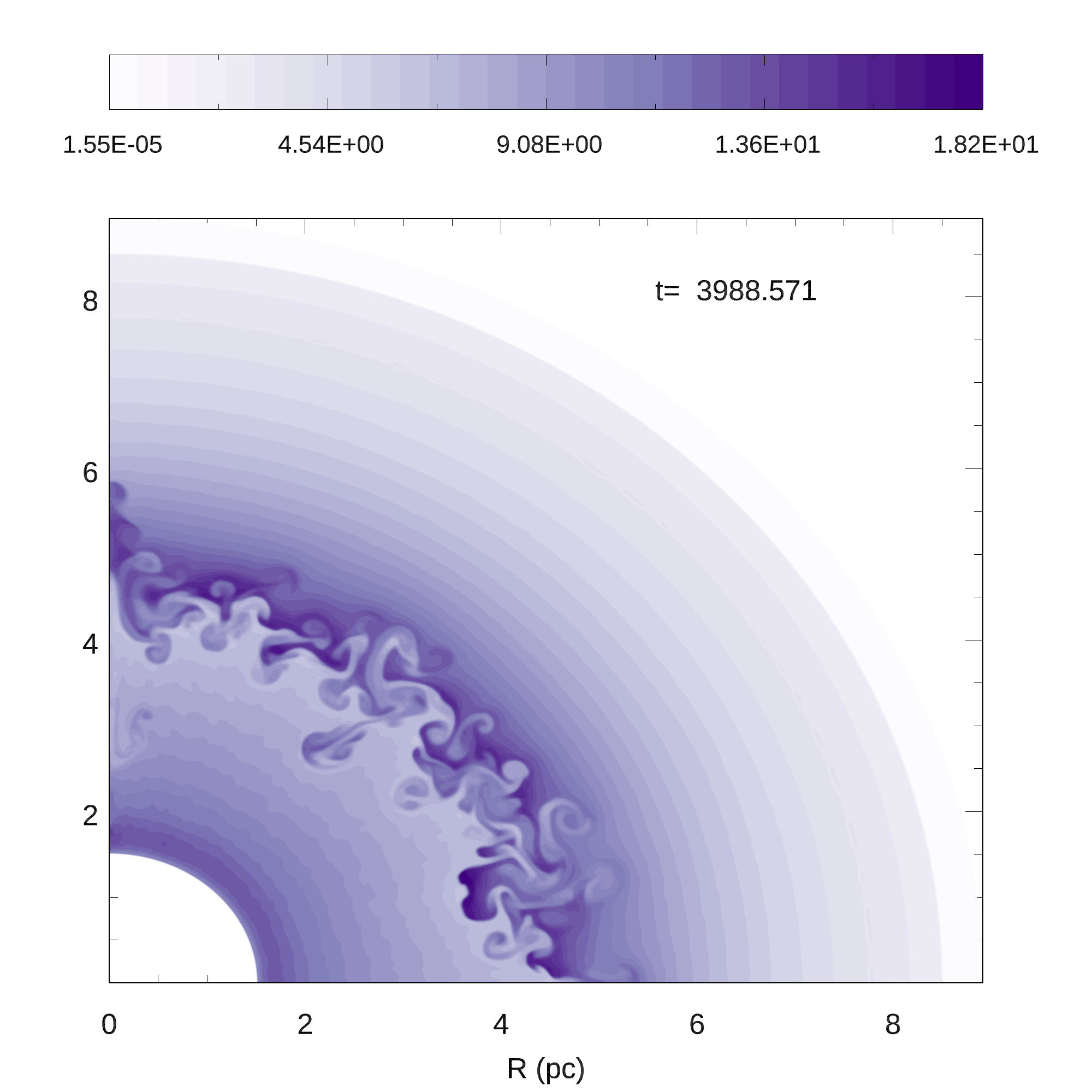}
\caption{\footnotesize A filled contour plot of the gas temperature (in keV) within the remnant at an
  age of 3988.6 years, from the 2-dimensional simulation described in
  the text. The contact discontinuity (CD) is found to be unstable,
  and R-T fingers topped by K-H caps are clearly seen around the
  CD. The effect of the instability is to spread the CD over a
  wide region, over 10\% of the remnant radius. It leads to a considerable variation of temperatures
  in the region of the CD, as shown.}
\label{fig:temp2d}
\end{center}
\end{figure}

In the spherically symmetric simulations, the CD is a sharp surface that cleanly separates the shocked ejecta from the
shocked CSM. In reality however the decelerating CD is unstable to the
Rayleigh-Taylor (R-T) instability \citep{cbe92}, leading to the growth
of R-T `fingers' extending out from the shocked ejecta into the
shocked CSM. Shearing between the fingers and the surrounding fluid
gives rise to the Kelvin-Helmholtz instability, and `mushroom caps'
can be seen atop the R-T fingers. The growth of R-T instability in
SNRs of Type Ia was studied by \citet{vvd00} for an exponential ejecta
profile. In order to study the multi-dimensional evolution and
kinematics, we have run a two-dimensional (2D) simulation using
approximately the same parameters as the 1-dimensional run. The simulation was
run with 500 radial and 500 angular zones on an expanding grid. As
expected, the CD was found to be unstable to the growth of R-T
instability. In Figure \ref{fig:temp2d} we show a frame from the 2D
simulation at approximately the same time as the 1D run above. The
image shows the temperature distribution across the grid. The
temperature profile follows more or less the 1D profile in Figure
\ref{fig:temp}, except at the contact discontinuity, where the turbulence
due to the R-T and K-H instabilities results in considerable
temperature variation around the CD, due to the presence of the R-T fingers that mix shocked ejecta with the shocked CSM. The effect of the temperature
(and density) variations is to effectively widen the CD, which is
now no longer a sharp spherical surface but has a width of around 10-15\% of the remnant radius (0.9- 1.4pc), centered at about 4.5 pc.

\section{SPI Analysis}
\label{section:analysis}

\subsection{Observations and Data Reduction}
\label{section:observationsreduction}

\begin{figure}[htbp]
\begin{center}
\includegraphics[width=0.4\textwidth]{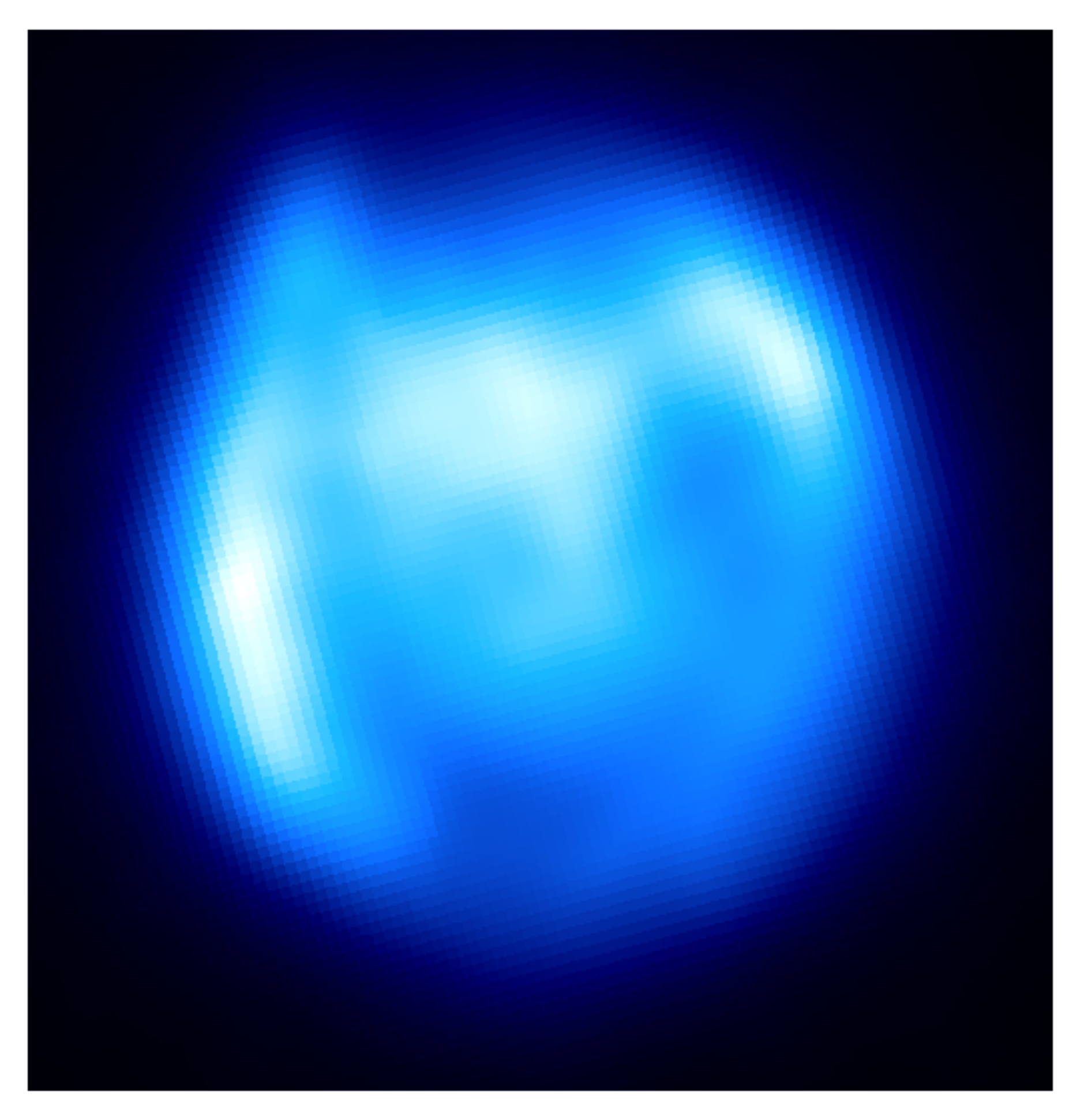}
\caption{\footnotesize Combined  EPIC-pn and MOS exposure-corrected image in the 0.2-8.0 keV band.}
\label{fig:xmmimage}
\end{center}
\end{figure}

We use the \xmms\ EPIC observation 0201840101 from 2003 December, the longest available observation of \dem. The exposure-corrected image is shown in Figure \ref{fig:xmmimage}.
Our SPI analysis requires only a filtered photon event list and exposure map for each detector (MOS1, MOS2, and pn). These were created using SAS 16.0, primarily the tasks {\em emchain} and {\em epchain}. To exclude non-X-ray events, the event lists were filtered to include only photon event patterns 0-12. They were filtered further to exclude events with energies outside of the 0.2-10 keV range.  Periods affected by soft proton flares were removed. The resulting number of events and exposure for each detector are shown in Table \ref{table:obs}. For the SPI analysis, only events within a $150''\times150''$ box centered on \dem\ were included. \dem\ is approximately circular with a radius of $\sim$45$''$.

\begin{table}
\begin{center}
\caption{XMM-EPIC Observation 0201840101}
\label{table:obs}
\begin{tabular}{ccc}
  \hline
  Detector & Events & Net Exposure (ks) \\
  \hline
  MOS1 & 175100 & 58.1 \\
  MOS2 & 357621 & 58.3 \\
  pn & 1034574 & 58.3 \\
  \hline
\end{tabular}
\end{center}
\end{table}

\subsection{Smoothed Particle Inference}
\label{section:SPI}
Smoothed Particle Inference \citep{Peterson2007} owes its flexibility to its modeling of the gas as a collection of independent `smoothed particles,' or \blobs\, of gas. Emission from all blobs is combined to represent the net X-ray emission from the SNR. Each blob has its own spectral and spatial model, typically including the gas temperature, abundances, spectral normalization, Gaussian width, and spatial position. Multiple \blobs\ can occupy the same line of sight, or even the same space, providing the ability to model a multiphase gas. While the \blobs\ themselves are spherically symmetric, no particular morphology or symmetry is assumed in their arrangement. The size and position of each \blob\ can be adjusted independently, which allows the multi-blob model to reproduce any arbitrary distribution (within limits imposed by the quality of the X-ray observation). This method is quite powerful, as it becomes a relatively straightforward task to characterize the distributions of any number of gas properties from the posterior distributions of the relevant blob parameters. SPI was specifically developed for the analysis of \xmm\ RGS and EPIC observations \citep{Peterson2007}, and has been previously applied to observations of galaxy clusters \citep{Andersson2007,Andersson2009,Frank2013}.

\subsection{Fitting Procedure}
SPI uses a Markov Chain Monte Carlo (MCMC) process to forward-fold the blob model and predict detector positions and energies for each photon. In this way SPI utilizes all available spectral and spatial information without imposing artificial restrictions on either the spatial or spectral distribution of the gas emission. A set of emitted photons is simulated for each blob, according to each blob's spectral and spatial model. These are then propagated through the instrument's response functions, including mirrors, gratings, and detector responses as necessary. Given the energy and location of each of these model photons, the probability of its detection is calculated for the given instrument response, as is its predicted location on the detector and the measured energy. The result is a set of model X-ray events which is compared to the observed events by binning both on the same three-dimensional grid (two spatial dimensions and energy). The two are then compared by calculating a two-sample likelihood statistic that quantifies the difference between the model and observed datasets. A modified Metropolis-Hastings algorithm \citep{Metropolis1953,Hastings1970} is used to choose a new set of blob parameters and the process is repeated. For more details of the SPI implementation, see \citet{Peterson2007}. After a number of iterations, the MCMC chain converges and the posterior parameter distributions become stable. After convergence, the fit typically has reduced $\chi^2$ values of 1.0-1.5. From this point onwards, each iteration returns a model that is statistically consistent with the data. 

The number of \blobs\ is fixed and chosen based on the spectral and spatial complexity of the emission, as well as the number of photons available in the data.  Previous work with galaxy clusters has suggested that the optimal number of \blobs\ is such that the number of model photons per \blob\ should be at least $10^4$ \citep{Frank2013}; this number can be increased (and computation time decreased) by reducing the number of \blobs. However, this criterion must be balanced with the need to have enough \blobs\ to adequately fit the more complex emission of SNRs. For our \dem\ analysis, 50 \blobs\ are used, resulting in $\sim$$3\times10^4$ photons per \blob. 

\begin{figure*}[h!tbp]
\begin{center}
\includegraphics[width=\textwidth]{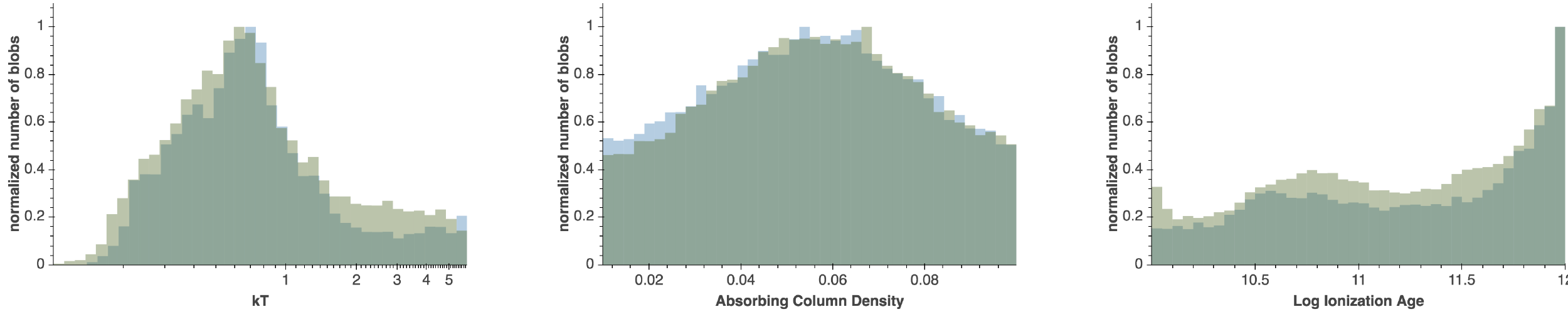}
\caption{\footnotesize Example comparisons of posterior distributions for SPI fits using 50 (blue) and 100 (green) \blobs.}
\label{fig:numblobcomparison}
\end{center}
\end{figure*}

\begin{figure}[htbp]
\includegraphics[width=0.5\textwidth]{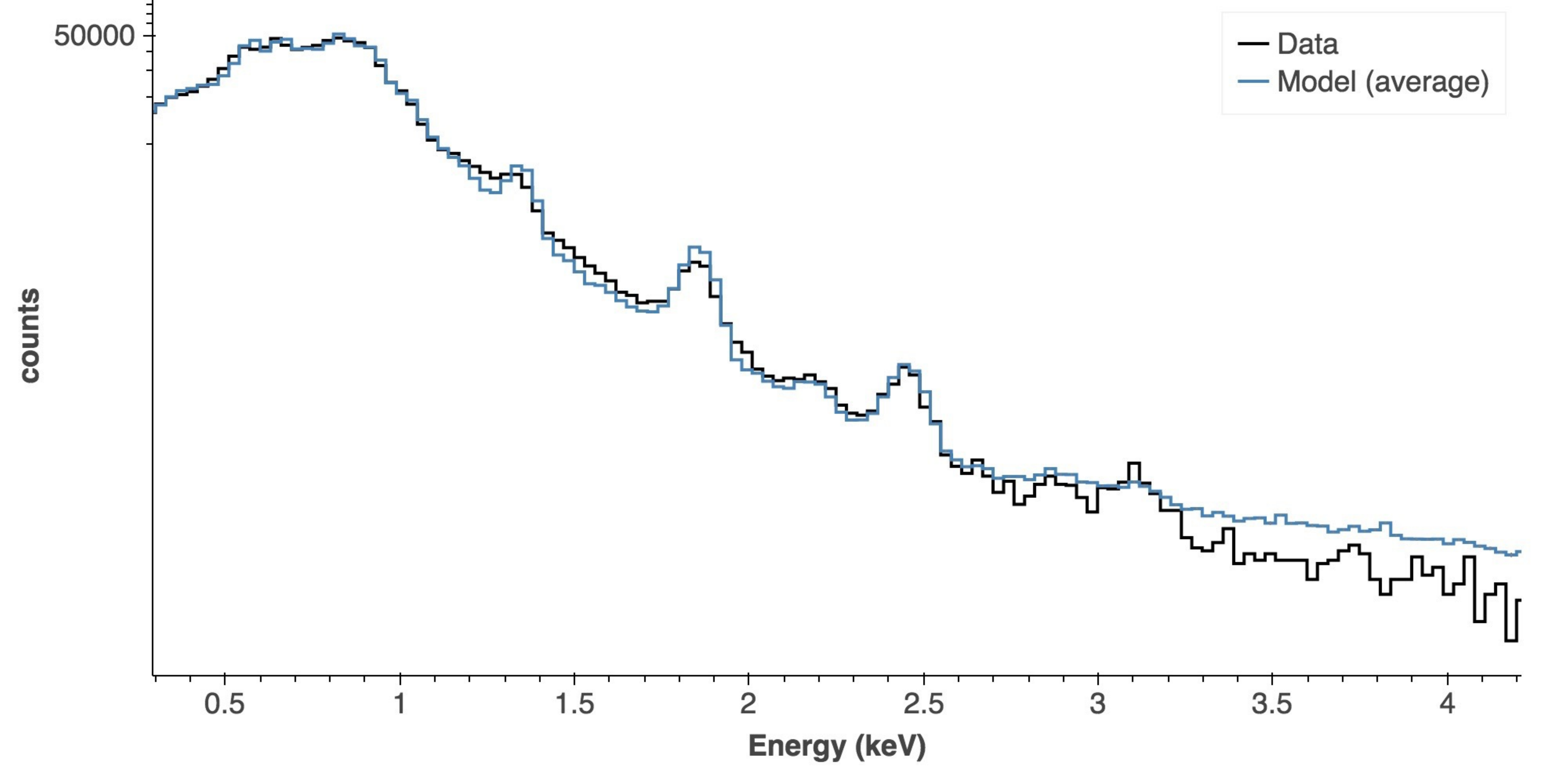}
\caption{\footnotesize Observed spectrum (black) and SPI model spectrum (blue).}
\label{fig:spectrum}
\end{figure}

The SPI fit of \dem\ results in a reduced $\chi^2=1.15$. We ensure that 50 \blobs\ provides sufficient accuracy by also testing with 100 \blobs\ and comparing the posterior distributions. Examples for temperature, absorbing column density, and ionization age are shown in Figure \ref{fig:numblobcomparison}. It is expected that the posteriors will not be identical, but they are extremely similar, indicating no improvement or significant change in the overall results when going from 50 to 100 \blobs. The reduced $\chi^2$ is also $\sim$1.15 for both the 50 and 100 blob models, and the medians, modes, and standard deviations of the parameter posteriors are nearly identical. The model spectrum also reproduces all the main features of the observed spectrum (Figure \ref{fig:spectrum}).

\subsection{Model}
\label{section:model}
With SPI, a combination of spectral and spatial models may be chosen to suit the particular astrophysical context, in this case an absorbed thermal plasma which is likely not in ionizational equilibrium. We also include model components to simultaneously fit the X-ray background. The model components are described in more detail below. The model is applied independently to each \blob; however some model parameters may be global (the same for all \blobs) and some are frozen. The spectral normalizations of each model component are always free parameters. All prior distributions are flat. The end points of the priors are chosen via an iterative process; initial guesses are made based on typical physical conditions in SNRs and anything known from previous X-ray analyses. We then test the model by performing an SPI fit and investigating the posterior distributions. The priors are adjusted if necessary and the MCMC restarted, to ensure that the priors span the entire relevant parameter space. For example, if the fit places an excess of \blobs\ at the upper limit of the prior (the largest allowed values), this upper limit is increased to allow exploration of higher values. Similarly, the limits are decreased if it is clear the parameter space spanned by the prior is unnecessarily wide, as this improves the efficiency of the MCMC.   

\subsubsection{Thermal Supernova Remnant Emission}
\label{section:thermalmodel}

The X-ray emission from \dem\ results from shocked plasma. In the case of DEML71, the emission is known to be thermal (H03, vdH03). However the gas may not necessarily be in ionization equilibrium, meaning that the ionization state of the gas may not be consistent with the post-shock temperature. Therefore we use the thermal, non-equilibrium ionization (NEI) plasma model {\em vpshock} \citep{Borkowski2001}. The free spectral parameters for each \blob, summarized in Table \ref{table:thermalpriors}, include temperature and ionization age (n$_e$t), as well as the abundances of O, Ne, Mg, Si, S, and Fe \citep[relative to solar abundances from][]{Anders1989}. All other abundances are fixed at typical LMC values of 0.5 solar \citep{Russell1992}. Free parameters may be different for every \blob, unless marked as global in Table \ref{table:thermalpriors}.  Note that temperature and ionization age are chosen from a logarithmically uniform prior. Volume emission measures can be easily derived directly from the spectral normalization of each \blob. Galactic absorption is included via the {\em phabs} absorption model, and is fixed to the \citet{Dickey1990} value of n$_H=5.7\times10^{20}$\nhunit, as it is not expected to vary substantially across such a small object. Absorbing material within the LMC is accounted for with a {\em zvphabs} model, with abundances set to average LMC values of 0.5 solar. This column density is tied for all model components, but is allowed to vary from \blob\ to \blob, to allow for the possibility of small spatial variations. Several of the background components also include Galactic and LMC absorption. The column densities for these components are tied to those associated with the {\em vpshock} component. Spatially, the {\em vpshock} emission is modeled as coming from a spherical Gaussian distribution (the `blob'). The central coordinates in the plane of the sky are allowed to vary independently for each blob anywhere within the $150''\times 150''$ box centered on \dem\, while the logarithmic prior for the Gaussian width of each blob is ln 2\farcs5$ <\sigma< $ ln $ 75''$ (thus the minimum diameter is $\sim5''$).

\begin{table}
\begin{center}
\caption{Model Priors}
\label{table:thermalpriors}
\begin{tabular}{cccc}
  \hline
  Parameter & Minimum & Maximum & Global \\
  \hline
  log kT (keV) & 0.08 & 7.0 & No\\
  n$_{H,LMC}$ (10$^{20}$\nhunit) & 1.0 & 10.0 & Yes \\
  log n$_e$t (cm$^{-3}$s) & 10$^{10}$ & 5$\times$10$^{12}$ & No \\
  O/O$_{\odot}$ & 0.01 & 1.0 & No \\
  Ne/Ne$_{\odot}$ & 0.1 & 1.5 & No \\
  Mg/Mg$_{\odot}$ & 0.1 & 1.5 & No \\
  Si/Si$_{\odot}$ & 0.01 & 3.0 & No \\
  S/S$_{\odot}$ & 0.01 & 3.0 & No\\
  Fe/Fe$_{\odot}$ & 0.01 & 3.0 & No \\
  \hline
\end{tabular}
\end{center}
\end{table}

\subsubsection{X-ray Background}
\label{section:backgroundmodel}
The X-ray background is a result of several sources of emission, including instrumental, cosmic, and Galactic. Each is accounted for with an additional model component. Background model components are fixed and the same for all \blobs\, except the spectral normalizations.

The EPIC instrumental background consists of soft protons, internal line emission, and electronic noise, and is accounted for with a custom model which includes all three components, as used in \citet{Frank2013}. The model is described in detail in \citet{Andersson2007}.

The Galactic X-ray background is soft, with energies $\lesssim1$ keV, and consists of several thermal components with $kT<0.5$ keV \citep{Lumb2002,Kuntz2000,Snowden2008}. We use two model components, one to account for the Local Hot Bubble (LHB) and one for slightly warmer emission from the Galactic Halo. Both employ thermal MEKAL spectral models, uniform across the field of view. The temperatures of the LHB and Halo are set to kT$=0.1$ keV and $0.25$ keV, respectively \citep{Snowden2008}. The LHB is unabsorbed, while Galactic absorption is applied to the Halo emission.

The final background component results from unresolved cosmic sources, mainly AGN, and contributes primarily at energies $\gtrsim1$ keV. These are modeled as a powerlaw, with $\Gamma=1.47$, that is spatially uniform. The powerlaw is absorbed by both Galactic absorption and absorption within the LMC.

\subsection{Post-SPI Analysis}
\label{section:postprocessing}
After a successful SPI fit, information about the SNR can be derived directly from the posterior parameter distributions, which include all \blobs\ from every iteration after convergence. Summary statistics, such as the medians and standard deviations, are the simplest to derive. Medians are similar to the averages that are measured with more traditional methods which extract a single spectrum from the entire SNR, or from several lines of sight through the SNR, and fit to a simple model. Standard deviations of each parameter can also be calculated, and are one of the advantages in using SPI, as they provide a straightforward measurement of how homogeneous the gas conditions are. 

Taking a step further, it can be even more informative to investigate the parameter distributions directly. SPI allows the construction of continuous parameter distributions (the posteriors), the shapes of which contain useful information on the gas composition and state. Distributions can be constructed from the set of \blobs\ by essentially creating histograms of the parameters of interest. 

In addition, because each \blob\ has a location and size, it is possible to create maps of any of the parameters. 
Maps are created by selecting a region containing the SNR and dividing it into spatial bins of a specified size (typically 1 to 5 arcsec). 
For each spatial bin, the contributions of each \blob\ are combined to determine the value. A number of different methods can be used to combine the \blobs\ within a spatial bin, depending on the purpose. The most common is to take the median value (e.g. the median of the blob temperatures), but it is also possible to use other quantities, including the maximum, minimum, sum, or standard deviation. The standard deviation is of particular interest, as it provides a means of mapping the level of homogeneity across the remnant. Note that these maps are all fundamentally different from X-ray images (including narrow-band images) or equivalent width images, which show only the intensity of related emission rather than measured values of the parameter. Narrow band and equivalent width images are fast methods for mapping the location of different elements or relative temperatures but do not measure the differences in abundances or other properties across the remnant. 

Associated statistical uncertainties can also be derived, as the probability density for each parameter is represented by the collection of models from all individual iterations. The uncertainty of a value $F$, $\delta_F$, can thus be characterized as the standard deviation of the distribution of $F$ over converged iterations,
\begin{equation}
\label{eqn:uncertainty}
\delta_F = \frac{1}{N_i}\sum_{i}^{N_i}[F_i-\overline{F}]^2,
\end{equation}
where $N_i$ is the number of iterations after convergence, the sum is over these iterations, and $F$ is the measurement of interest, a function of the \blob\ parameters, for example the overall median or mean temperature. Systematic uncertainties are much more difficult to estimate. The shape of the posterior distributions can provide some information. The posterior shape is influenced by a several factors in addition to the true parameter distribution. First is the shape of the prior; the prior distributions are all uniform, and thus the further the posterior deviates from uniform, the more confident we can be that the SPI fitting has successfully constrained the distribution. The converse, however, is not true; a uniform posterior may be the result of the true distribution also being uniform, and in this case it is not distinguishable from the prior. Second, there is some broadening of the posterior distribution due to the imperfect quality of the data and the inherently incomplete spectroscopic information about the gas properties that can be derived from atomic transitions. This broadening will be smaller for higher quality data and can be minimized by choosing an optimal number of blobs \citep[][]{Frank2013}.
Because of the difficulty in assessing the details of these systematic uncertainties, small scale features of the distributions should be interpreted with caution, but the basic shape of the distributions are sufficiently robust to be informative.

The raw summary statistics, distributions, and maps, for which all blobs are treated equally, are useful for exploring the results and identifying minor but important components of the SNR gas. However, it is usually more meaningful to first weight each \blob\ by its emission measure (EM), to account for the different sizes and densities and thus better represent the true contribution of each \blob\ to the overall SNR. 

\begin{table*}[htbp]
\centering
\caption{EM-weighted summary statistics.} 
\label{table:emstats}
\begin{tabular}{lcccccc}
\hline
 & Median & Mode & Standard Deviation \\
n$_{H,LMC}$ ($10^{20}$ cm$^{-2}$) & 6.7$\pm$0.9   & 9.0$\pm$1.7   & 2.4$\pm$0.3   \\
kT (keV)      					& 0.24$\pm$0.04    & 0.15$\pm$0.02    & 0.43$\pm$0.07    \\
log $n_et$ (cm$^{-3}s$) & 12.4$\pm$0.1    & 12.7$\pm$0.2    & 0.6$\pm$0.1    \\
O/O$_{\odot}$   & 0.19$\pm$0.04    & 0.02$\pm$0.08    & 0.27$\pm$0.02    \\
Ne/Ne$_{\odot}$  & 0.45$\pm$0.12    & 0.11$\pm$0.30    & 0.35$\pm$0.05    \\
Mg/Mg$_{\odot}$ & 0.75$\pm$0.14    & 0.95$\pm$0.36    & 0.37$\pm$0.05    \\
Si/Si$_{\odot}$  & 1.19$\pm$0.30    & 0.40$\pm$0.77    & 0.79$\pm$ 0.10    \\
S/S$_{\odot}$ & 1.43$\pm$0.27    & 1.54$\pm$0.74    & 0.77$\pm$0.09    \\
Fe/Fe$_{\odot}$  & 0.74$\pm$0.31    & 0.16$\pm$0.85    & 0.83$\pm$0.12    \\
\hline
\end{tabular}
\end{table*}

SPI also provides the capability to explore any subset of the parameter space by selecting a set of \blobs\ based on any combination of parameters and investigating the summary statistics, distributions, and maps of only that subset. For example, it is possible to select only \blobs\ in a particular temperature or ionization age range (or both) and then investigate the corresponding abundance distributions or maps. Filtering in this way has huge advantages for isolating and measuring the properties of specific physical components in the SNR, as it eliminates confusion from unrelated material that may be along the same line of site. This is one of the most powerful and unique capabilities of SPI.


\section{Application of SPI to DEM L71}
\label{section:results}
We detail here the results of the application of the SPI analysis, as outlined in Section \ref{section:analysis}, to the SNR \dem.

\subsection{Summary Statistics}
\label{section:stats}

The EM-weighted summary statistics are shown in Table \ref{table:emstats}. The volume emission measure for the entire remnant is $EM = 5.53 (\pm0.59) \times10^{59}$ cm$^{-3}$. These simple quantities provide a very limited view of the actual parameter distributions, shown in Figure \ref{fig:dists2}, but they are useful for comparison with more standard spectral fitting methods. 

In most cases, large standard deviations and/or substantial differences between the median and mode indicate the distribution is complex enough that a simple mean or median is a poor descriptor. For example, the median EM-weighted temperature is 0.54 keV, but a very different mode (0.22 keV) indicates the actual distribution is more complicated. Similarly, O, Ne, Mg, and Fe EM-weighted median abundances are all roughly similar to typical LMC values ($\sim$0.5 solar), while Si and S are higher. If this was the only information available, it might be concluded that O, Ne,  Mg, and Fe are all from shocked CSM, while the Si and S is shocked ejecta. However, in the cases of O, Ne, and Fe, the other summary statistics suggest the situation is not so simple; the modes are all very low, implying there are regions with \dem\ that are largely devoid of O, Ne, or Fe, and the standard deviation of Fe is large enough to indicate the presence of not only gas with very low Fe, but also very high Fe. The Si mode of 0.40 is more indicative of shocked CSM than ejecta.  

\subsection{Parameter Distributions}
\label{section:distributions}
\begin{figure*}[htbp]
\begin{center}
\includegraphics[width=\textwidth]{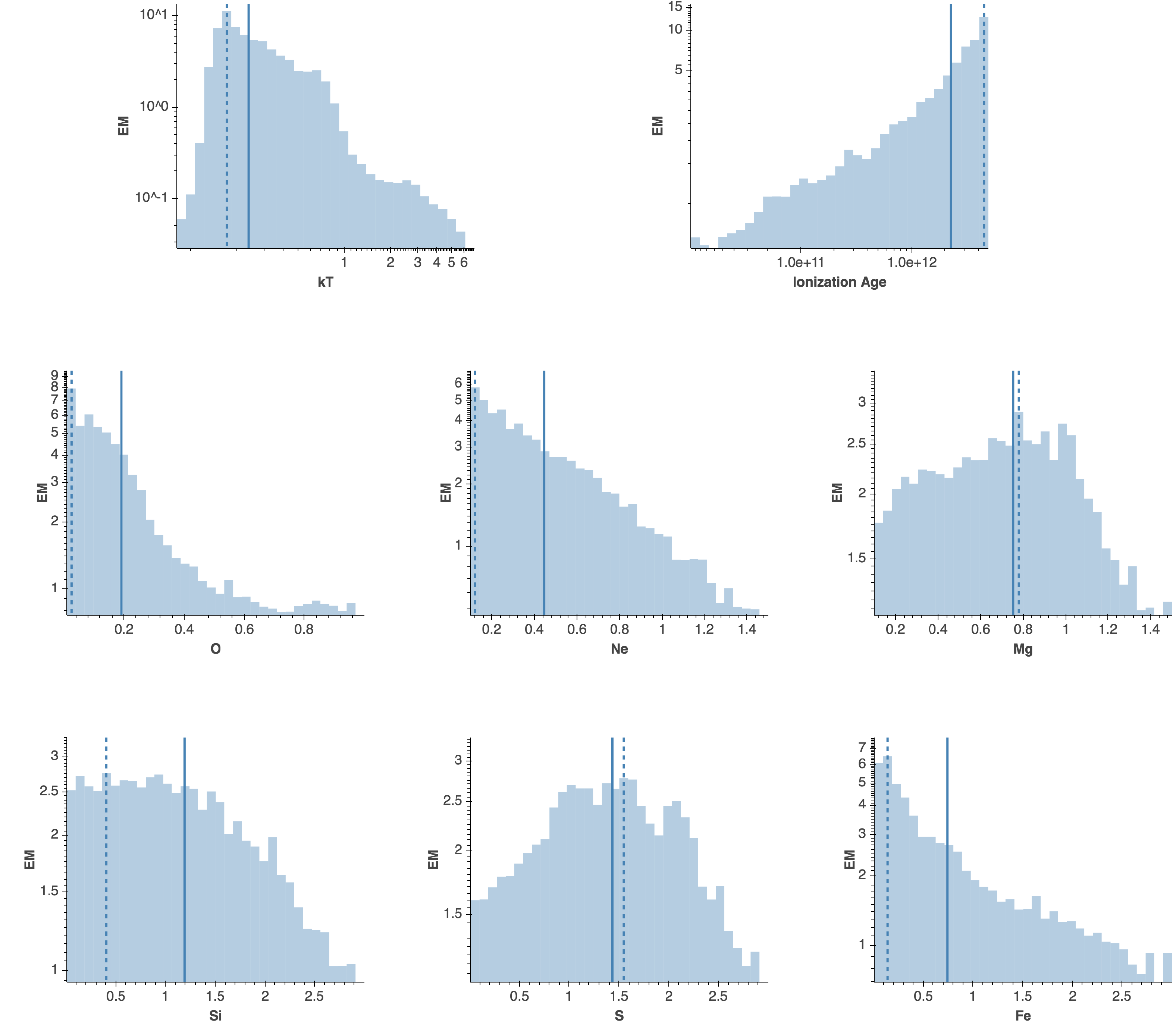}
\caption{\footnotesize Emission-measure-weighted parameter posterior distributions. Units of abundances are relative to solar. kT is units of keV, ionization age in cm$^{-3}$s. Emission measure is given in units of 10$^{62}$ cm$^{-3}$. The modes and medians are shown as dashed and solid lines, respectively. }
\label{fig:dists2}
\end{center}
\end{figure*}

Emission-measure-weighted posterior distributions are shown in Figure \ref{fig:dists2}. Typically some \blobs\ are present throughout the entire allowed parameter space. 
The temperature distribution reveals most gas has temperatures $\lesssim$$1$ keV, but some does exist at higher temperatures with substantially lower EM. Ionization age tends to be above 10$^{12}$, indicating much, but not all, of the gas is in ionization equilibrium\footnote{Ordinarily the fact that the ionization age peaks at the highest allowed values would suggest the prior should be extended. However, in this case all higher values would be in ionization equilibrium, and thus increasing the range is not justified.}. O shows a clear peak at $\lesssim$0.3, in line with the existence of some O-deficient gas, while the majority of O emission comes from the shocked CSM. 
The Ne distribution is similar, but with a more gradual decline at higher values. 
Mg peaks near typical LMC values of $\sim1$ and drops off sharply above solar. 
Si also tends to have abundances $\lesssim1$, but with a substantial contribution from higher abundance material. S has a rather wide distribution, and likely is not well-constrained, but the distribution suggests that S abundances $>2.5$ are unlikely. Most gas in \dem\ clearly has low Fe, $\lesssim1$, but some exists with higher Fe abundances as well, up to $\sim$$2.5$.

\subsection{Maps}
\label{section:maps}
\begin{figure*}[htbp]
\begin{center}
\includegraphics[width=\textwidth]{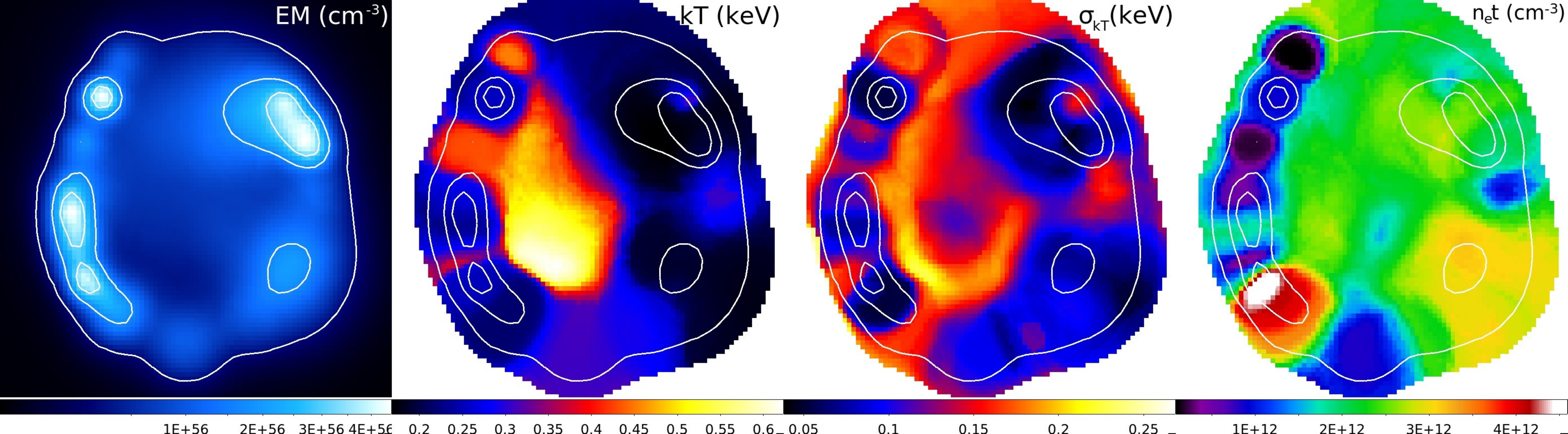}
\caption{\footnotesize From left to right, maps of the emission measure (cm$^{-3}$, square root scale), EM-weighted temperature (keV), EM-weighted standard deviation of the temperature, and EM-weighted $n_et$ (cm$^{-3}$s). The values in all except the EM map are set to zero where the EM (left) was less than 0.05\% of the mean, to avoid noise due to poor statistics on the outer edges.
        Spatial bins are 1 arcsec$^2$. Contours are taken from the emission measure map.}
\label{fig:maps1}
\end{center}
\end{figure*}

\begin{figure*}[htbp]
\begin{center}
\includegraphics[width=0.75\textwidth]{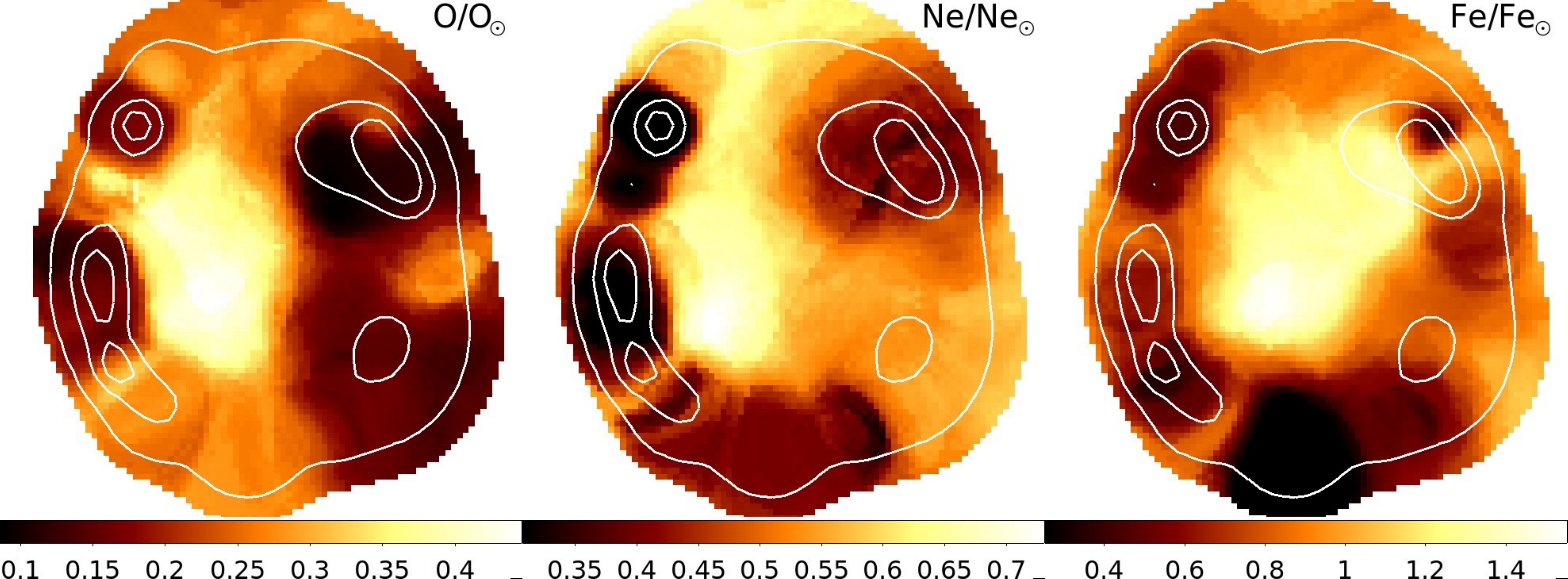}
\caption{\footnotesize Emission-measure-weighted abundance maps of O, Ne, and Fe. The values in all except the EM map are set to zero where the EM (Figure \ref{fig:maps1}, left) was less than 0.05\% of the mean, to avoid noise due to poor statistics on the outer edges. Spatial bins are 1 arcsec$^2$. Contours are taken from the emission measure map (Figure \ref{fig:maps1}, left).}
\label{fig:maps2}
\end{center}
\end{figure*}

EM-weighted parameter maps are shown in Figures \ref{fig:maps1} and \ref{fig:maps2}. Overall, the center of \dem\ is hotter, with lower EM, and larger Fe, Ne, and O. Apart from these general trends, the maps are all highly irregular. The outer rim is particularly clumpy, with azimuthal variations of at least a factor of two in all parameters.  

\subsection{Identifying Ejecta Emission} 
\label{section:subsets}

\begin{figure}[htbp]
\begin{center}
\includegraphics[width=\columnwidth]{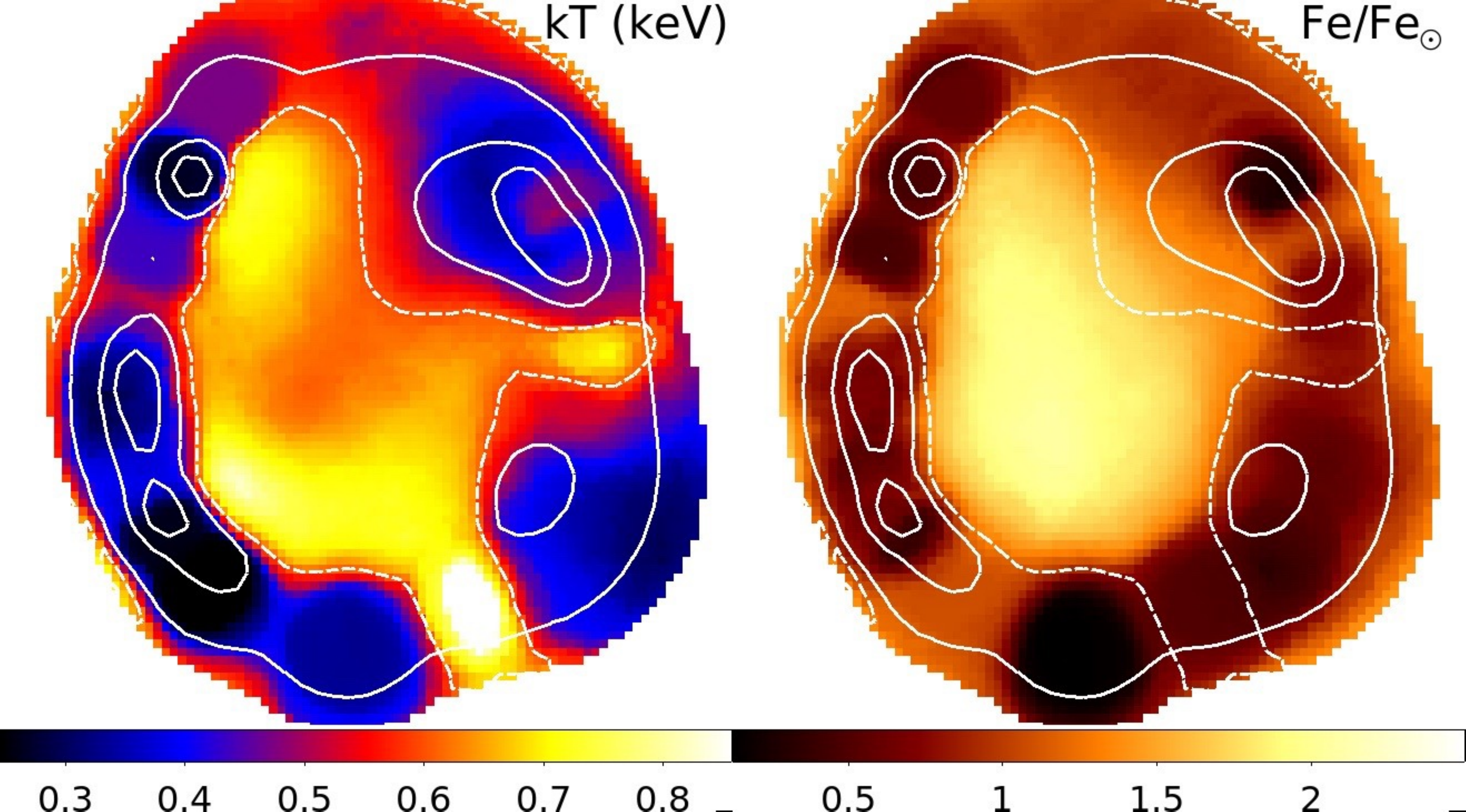}
\caption{\footnotesize Unweighted maps of kT (keV, left) and Fe (right). Values are set to zero where the EM was less than 0.05\% of the mean, to avoid noise due to poor statistics on the outer edges.
        Spatial bins are 1 arcsec$^2$. Solid contours are taken from the emission measure map (Figure \ref{fig:maps1}, left). Dashed contours representing kT=0.6 keV on the kT map (left)  are shown on both maps, demarcating the clearly warmer and higher Fe gas in the core.}
\label{fig:unweightedmaps}
\end{center}
\end{figure}

\begin{figure*}[htbp]
\begin{center}
\includegraphics[width=0.75\textwidth]{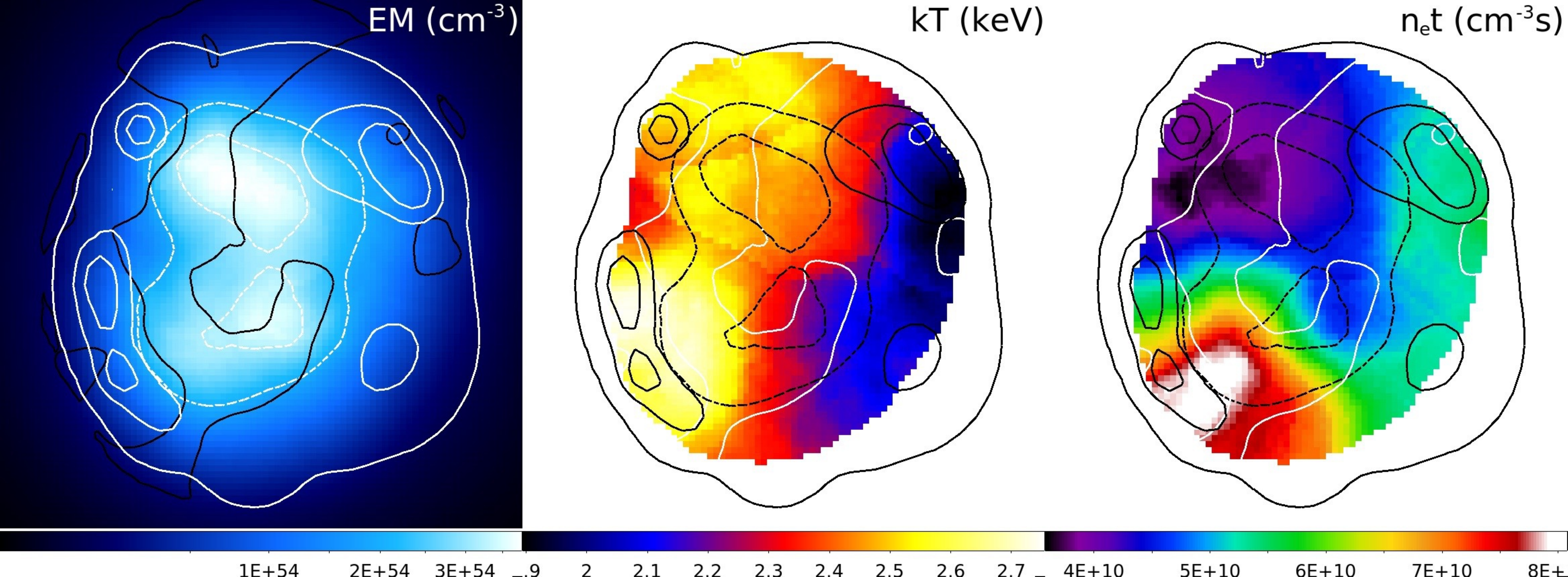}
\caption{\footnotesize From left to right, maps of the emission measure (cm$^{-3}$, square root scale), EM-weighted temperature (keV), and EM-weighted $n_et$ (cm$^{-3}$s) for only the ejecta emission, as defined in Section \ref{section:subsets}. The temperature and ionization age are set to zero where the EM (left) was less than 0.05\% of the mean, to avoid noise due to poor statistics on the outer edges.
        Spatial bins are 1 arcsec$^2$. Solid black contours are taken from the overall emission measure map (Figure \ref{fig:maps1}, left), and dashed black contours from the ejecta emission measure map (left, this figure). White contours represent $\sigma_{kT}$=0.14 keV from Figure \ref{fig:maps1} (center right), illustrating the hook-shaped region with large temperature variations (contour colors are inverted on the EM map).}
\label{fig:ejectamaps}
\end{center}
\end{figure*}

\begin{figure*}[htbp]
\begin{center}
\includegraphics[width=\textwidth]{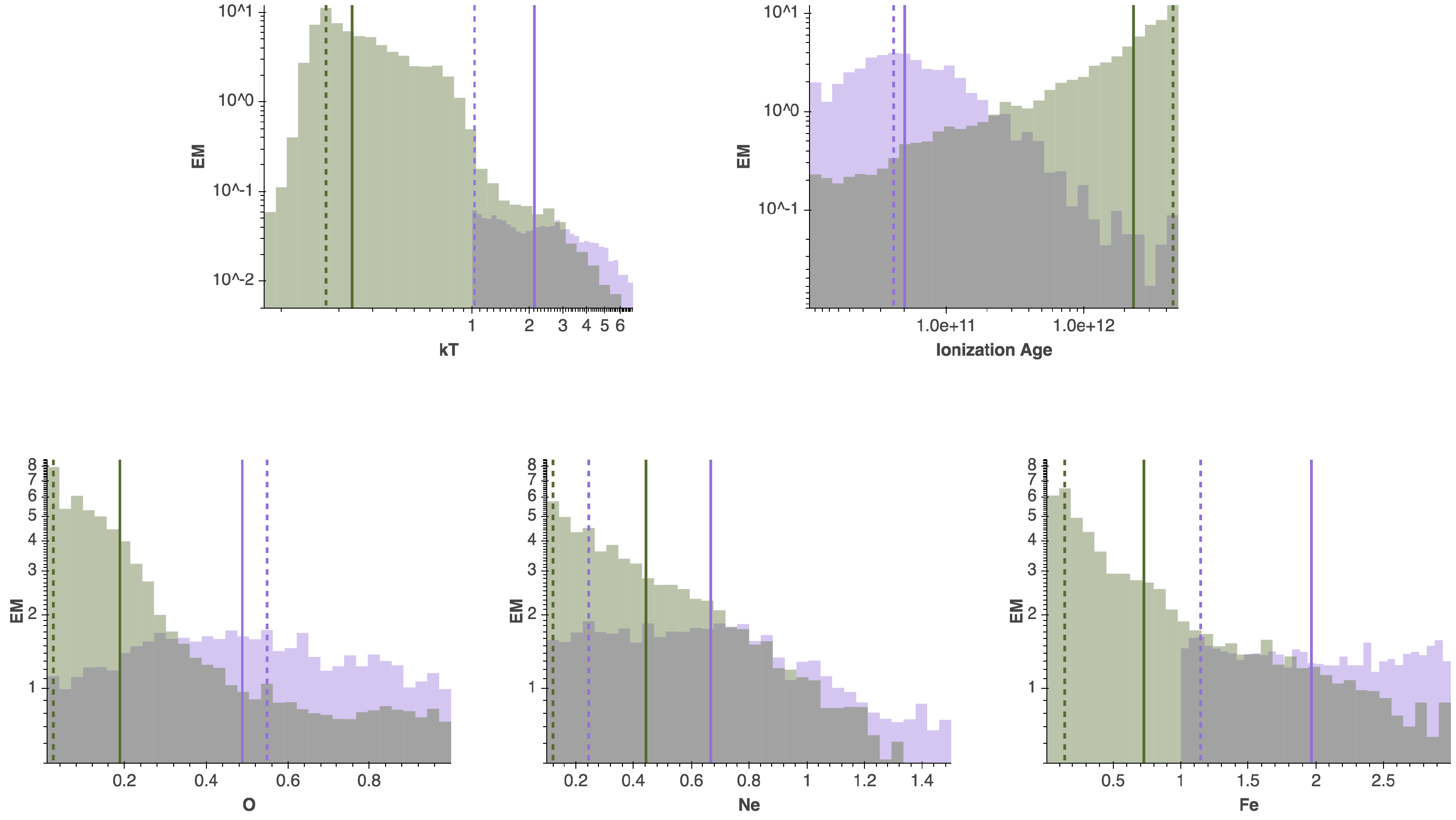}
\caption{\footnotesize EM-weighted temperature,  ionization age, O, Ne, and Fe distributions for the ejecta emission (purple) and the CSM emission (green). 
Units of abundances are relative to solar. EM is in units of $10^{62}$ cm$^{-3}$.
The modes and medians are shown as dashed and solid lines, respectively. The ejecta emission measures have been scaled up by a factor of 40 to aid visual comparison, with the exception of in the temperature distribution.}
\label{fig:ejectadists}
\end{center}
\end{figure*}

The substantially higher EM of the clumps in the outer rim implies that the EM-weighted distributions and maps (Figures \ref{fig:dists2}-\ref{fig:maps2}) are dominated by the outer rim material. However, the parameter distributions (Figure \ref{fig:dists2}) clearly show the presence of of some lower EM gas. 
Maps which are not weighted by EM, and therefore give equal weight to both high and low EM \blobs, can provide some insight on the properties of the lower EM gas. Unweighted temperature and Fe maps (Figure \ref{fig:unweightedmaps}) indicate that the lower EM core gas has both higher temperatures and high Fe. 
As demonstrated in the simulations (Figure \ref{fig:temp}), this is expected for emission from the shocked ejecta, which is also expected to have lower density and lower $n_et$. 
Therefore to investigate the core gas, we selected only \blobs\ with kT$>1$ keV and Fe$>$1, a combination which produces a clearly centralized spatial distribution (Figure \ref{fig:ejectamaps}, left). Selecting \blobs\ with particular characteristics to measure properties of the ejecta dramatically reduces contamination in these measurements from overlying CSM material. Reducing either the temperature or Fe thresholds results in the inclusion of significant amounts of the high EM gas in the outer rim, and thus will not represent the core emission, while the maps and parameter distributions with higher thresholds look nearly identical to those with kT$>1$ keV and Fe$>$1. In addition to being warmer with higher Fe, the core 
emission has distinctly lower ionization ages, $\lesssim 10^{12}$, and higher O, as can be seen in the associated parameter distributions (Figure \ref{fig:ejectadists}). The temperature and ionization age maps both show a gradient, with the hottest gas and highest ionization ages in the southeast and lowest in the west (temperature) and northeast (ionization age).

Our 2D simulations illustrate that the unstable CD, which is spread over a wide region due to R-T instabilities, has the largest temperature variations (Figure \ref{fig:temp2d}). Using SPI, we can measure and map the temperature variations, as seen in Figure \ref{fig:maps1} (center right). This is a map of the standard deviation of the temperature within each spatial bin, 
and clearly shows a wide hook-shaped structure in the outer core where the variations are largest. Both the width of this region, $\sim$1.7pc, and its distance from the center, $\sim$4.5pc, agree very well with the location and width of the CD from the simulations (Figure \ref{fig:temp2d}). 
Although the widths from the SPI map and the 2D simulations are both upper limits, and the hook shape itself cannot be reproduced in our spherically symmetric simulations, the agreement on the size and location of this region with maximum temperature variations allows it to be identified as the location of the CD and lends further support for the core emission inside this region to be arising from the SN ejecta. 

It is also clear from the simulations that the temperature in this region will be higher than in the rest of the ejecta, and indeed both the overall (Figure \ref{fig:maps1}) and ejecta (Figure \ref{fig:ejectamaps}) maps indicate that the highest temperatures are found there. 
\maggi\ used two temperature components to fit the ejecta, while \vdH\ used a shell like structure for the core region. Although the above discussion provides some understanding of the basis for these assumptions, we emphasize that neither of them is required when using the SPI technique. 
Any asymmetry in the shape of the CD region in SPI will be due to asymmetries in the ejecta or the medium into which the remnant was expanding, or due to projection effects, and are not captured in our simulations.

\section{Discussion}
\label{section:discussion}

Previous measurements of the physical properties of \dem\ rely on extracting X-ray spectra from a specified regions and fitting a NEI spectral model, with one to three components to represent the CSM and ejecta contributions. Ejecta properties are measured by designating one or more of the spectral components to be ejecta (e.g. by allowing higher abundances). \hughes, \rakowski, \vdH, and \maggi\ all take this approach using \xmm\ (\vdH\ and \maggi) and \chandra\ (\hughes\ and \rakowski) observations. 
Both \maggi\ and \vdH\ extract and fit an EPIC-pn spectrum from the entire source region. To investigate CSM vs ejecta properties, \vdH\ also extract and fit separate spectra for two regions, an outer annulus and a core region, while \maggi\ instead use a three-component model to fit their single spectrum (one CSM component and two ejecta). Rather than fit spectra that encompass the entirety of \dem, \hughes\ and \rakowski\ take advantage of the higher resolution of \chandra\ to extract spectra from several small regions. \hughes\ choose 4 small regions in the southeast to represent the inner and outer CSM and the inner and outer ejecta, fitting the spectra from each independently. \rakowski\ extract spectra from regions in the bright outer shell with 5 different azimuthal positions, representing the CSM, and carry out a joint fit to determine abundances. Together, these works represent the most common approaches to X-ray analysis of SNRs, and thus make a useful basis for comparison to assess the efficacy of SPI.

\begin{figure*}[htbp]
\begin{center}
\includegraphics[width=\textwidth]{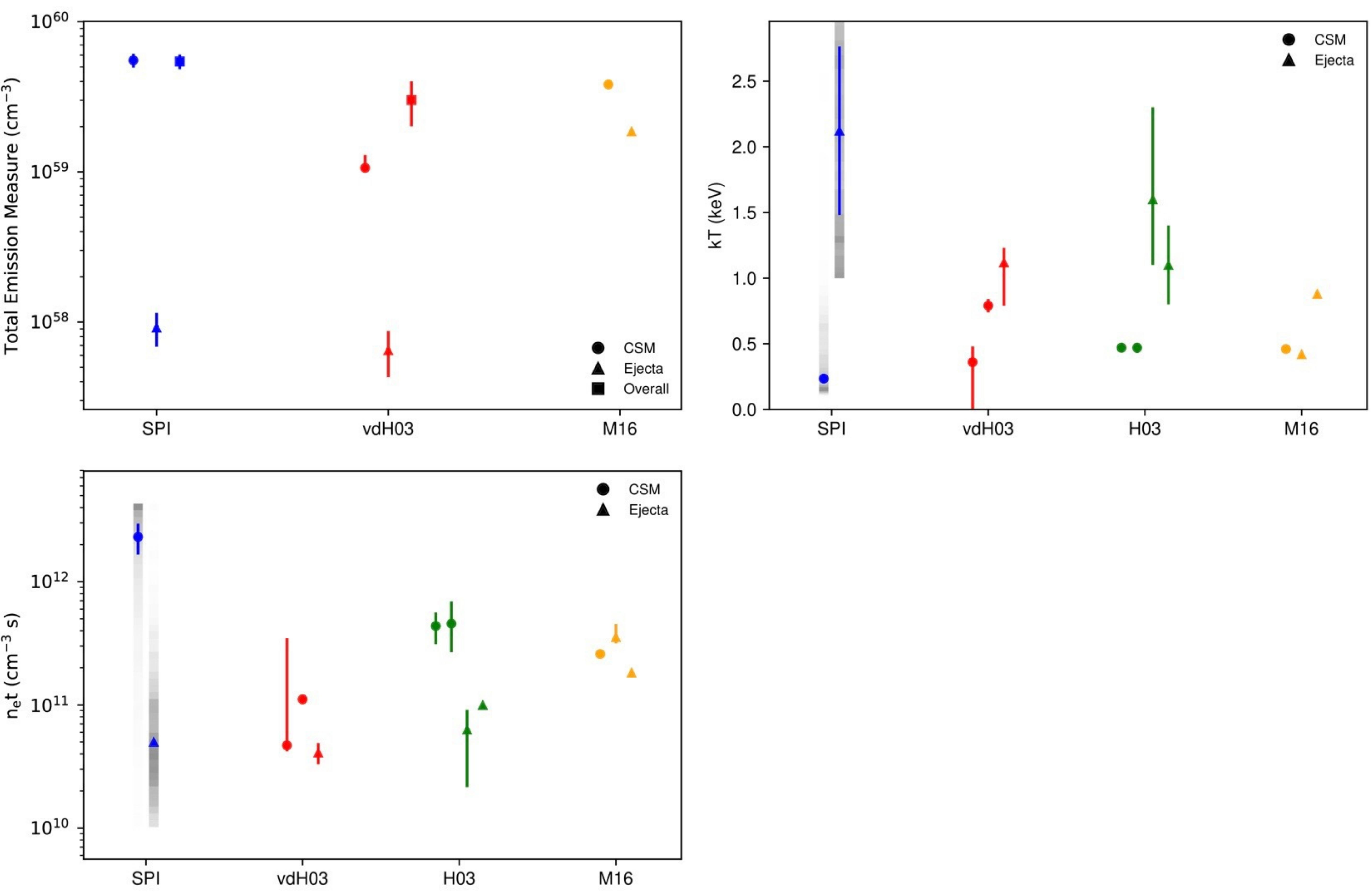}
\caption{\footnotesize Comparison of EM-weighted  medians from the SPI analysis in this work (blue) with values reported by \vdH\ (red), \hughes\ (green), \maggi\ (orange), and \rakowski\ (purple). Ejecta measurements are shown as triangles and the CSM as circles. Gray bars represent the posterior parameter distributions for the CSM and ejecta (such as those shown in Figure \ref{fig:ejectadists}), with darker corresponding to more EM (bars do not share EM scale).}
\label{fig:litcomparisons1}
\end{center}
\end{figure*}

\begin{figure*}[htbp]
\begin{center}
\includegraphics[width=\textwidth]{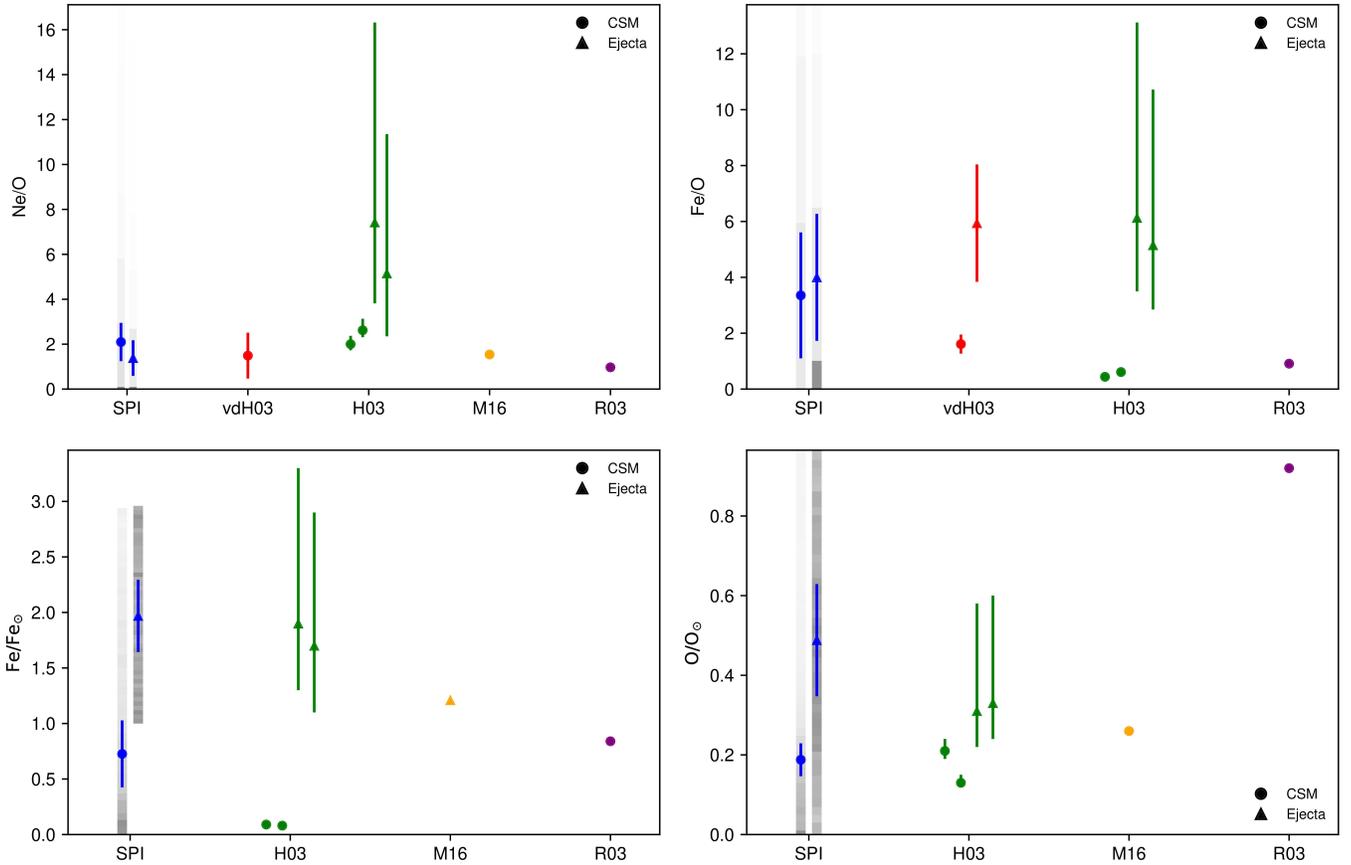}
\caption{\footnotesize Comparison of EM-weighted Ne, O, and Fe abundance medians from the SPI analysis in this work (blue) with values reported by \vdH\ (red), \hughes\ (green), \maggi\ (orange), and \rakowski\ (purple). Ejecta measurements are shown as triangles and the CSM as circles. Gray bars represent the posterior parameter distributions for the CSM and ejecta (such as those shown in Figure \ref{fig:ejectadists}), with darker corresponding to more EM (bars do not share EM scale). Ne/O CSM values from \vdH\ are taken from their fit to the total spectrum as they do not report this value for the CSM separately.}
\label{fig:litcomparisons2}
\end{center}
\end{figure*}

All of these analyses result in discrete measurements of the properties of interest. The SPI equivalents to these measurements are the EM-weighted medians as in Table \ref{table:emstats}. We therefore expect that these values should be similar to those reported by other authors, and indeed we find this is generally true. Figures \ref{fig:litcomparisons1} and \ref{fig:litcomparisons2} compare our medians with the equivalents reported in other works, both for the ejecta and for the CSM. In addition to the simple medians, we show shaded gray bars representing the EM-weighted posterior distributions as in Figure \ref{fig:ejectadists}. 
The overall picture is consistent: a bright shell of CSM surrounding a warmer core. The high EM shell is cool, kT $\sim0.3-0.5$ keV, with ionization ages of at least a few times $10^{11}$cm$^{-3}$s and subsolar O, Ne, Mg, and Fe abundances that suggest shocked CSM. The core has significantly lower EM and is warmer (kT $\sim1-2$ keV) with lower ionization ages and enhanced abundances. 
However, there are a number of differences between previously reported results and the results reported here that are informative. 

\maggi, \vdH, and this work all use \xmm\ EPIC-pn observations. The total emission measure obtained from our SPI analysis is 5.53($\pm0.59)\times$10$^{59}$cm$^{-3}$. This matches that from \maggi, 5.68$^{+0.07}_{-2.90}\times$10$^{59}$cm$^{-3}$. For their fit to the full EPIC-pn spectrum, \vdH\ find a total emission measure of 3.54($\pm1.05)\times$10$^{59}$cm$^{-3}$, which is somewhat smaller but still reasonably close. 
Separating the CSM and ejecta emission, the total EM of the ejecta found in this work agrees with that of \vdH, but the EM of the CSM is substantially lower in \vdH\ (see Figure \ref{fig:litcomparisons1}).  
The combined EM from their separate CSM and ejecta spectral fits is only 1.13$^{+0.23}_{-0.04}\times$10$^{59}$cm$^{-3}$, substantially lower than the total from their fit to the integrated spectrum. It is clear that such a model, where all the X-ray emission comes from two shells, ignores a substantial fraction of the remnant emission (everything outside the individual regions).
\maggi\ report a similar CSM EM, but much higher ejecta EM than either \vdH\ or this work. These differences are most likely due to the very different assumptions made during the spectral fits, both in the number of components and the abundances. . 
All of these measurements of the total EM are based on spectra from the same instrument (EPIC-pn) extracted from the same region (all of \dem). The differences found here illustrate the inherent uncertainties involved when assumptions are made about the spectral state of the emitting gas, especially when only a small fraction of the gas is being considered, something that is largely avoided with SPI. 

The majority of the X-ray emission is from the low temperature gas in the CSM, but there is clearly some contribution from gas with temperatures above 2 keV, as can be seen in the temperature distribution (Figure \ref{fig:dists2}). Our maps show this hot gas is mostly located in the core (Figures \ref{fig:maps1}, \ref{fig:unweightedmaps}, and \ref{fig:ejectamaps}). However, simple fits miss this low emission measure gas. 
This is a byproduct of only being able to measure discrete values that are similar to a weighted average over a region. While the low EM gas may influence these measured averages, resulting in the higher reported temperatures of the ejecta for example, the properties of this gas cannot generally be determined by standard spectral fitting methods, since it is very difficult to separate the emission from the more dominant (high EM) gas. Temperature is the clearest example, but the same effect applies to all parameters. Extremely deep observations can help, but our results clearly demonstrate that the physical state of the gas is very complex. Standard spectral analyses require a discrete number of spectral components, but observations which are of high enough quality to explore the full complexity of the gas cannot be adequately fit with a manageable number of these components. 

While the \xmm\ analyses use large spatial regions, resulting in averaging over any variations, the \chandra\ studies take the opposite approach, with a few small regions chosen for spectral analysis. This has the advantage of a simpler spectrum, as a smaller region will be more physically homogeneous and is therefore easier to fit, although the spectrum is still integrated along the line of sight. 

\hughes\ provide the only other measurement of Ne in the ejecta, and find a very high abundance (Figure \ref{fig:litcomparisons2}) compared to the CSM: 1.7 and 2.3 times solar for the inner and outer ejecta, respectively. We do not find this to be the case, instead finding LMC-like values that are similar to those in the CSM. The Ne distribution (Figure \ref{fig:ejectadists}) does indicate a very small amount of gas with Ne $>$ 1 in the ejecta, and Ne is highest in the region used by \hughes\ in the southeast (Figure \ref{fig:maps2}), but in neither case do we find a substantial amount of Ne gas with such a high abundance as indicated by \hughes. However, note that the uncertainties on the reported high Ne are large, likely a consequence of choosing a small region; there are relatively few counts in the spectrum.

In contrast to Ne, we find higher Fe in the CSM than either \hughes\ or \rakowski. This can be explained by the choice of the spectral extraction regions. As with all parameters, we find that Fe varies irregularly across \dem\ (Figure \ref{fig:maps2}). In choosing only a few small regions, \hughes\ and \rakowski\ are sampling different parts of this spatial distribution, and thus measure Fe values for the CSM that are different from both each other and this work. The region used by \hughes, in the southeast, is among the lowest Fe regions in \dem, and our Fe distribution (Figure \ref{fig:dists2}) does peak at very low Fe (the asymmetric shape is what causes the median to be higher). \rakowski, on the other hand, combines several regions from around the bright CSM shell and correspondingly measures a higher value for Fe, similar to the median Fe we report here. While this better matches our median Fe, it does not encompass the low Fe gas indicated by both our results and that of \hughes. These differences in abundance measurements highlight the selection effects resulting from the practice of choosing discrete regions. Because physical conditions vary widely and irregularly across the entire source, the results are highly dependent on choice of region and are unlikely to be representative of the object as a whole.

\section{Conclusions}
\label{section:conclusions}

SPI analysis of \dem\ reveals a more comprehensive and complicated picture than previously known, with irregular distributions of temperature, ionization age, and other properties. The cool outer shell of shocked CSM has a median temperature of $0.24$ keV but includes gas with temperatures that range from $\sim0.2$ to $\sim0.45$ keV. Ionization age of the CSM material varies widely around the outer rim but is high on average, indicating much of the CSM is in ionizational equilibrium. The CSM exhibits subsolar LMC-like abundances. The inner core is warmer with enhanced Fe (greater than solar), which supports a Type Ia origin for \dem. The core also reveals a somewhat higher, but still subsolar, abundance of O and Ne. We use the enhanced Fe and warmer temperatures to isolate the ejecta from the CSM and examine it in more detail. We find that in addition to typical temperatures of $\sim1-3$ keV, the ionization age of the core material is much lower than that of the CSM. Both the warmer temperatures and lower ionization age of the ejecta are in agreement with expectations from our simulations. We  also independently identify the contact discontinuity that separates the shocked CSM from the ejecta via a distinctive hook-shaped region where temperature variations are at a maximum, matching the R-T unstable region predicted BY our 2D simulations.

Overall, we find median values that generally agree well with what other studies have found; however, these are broad averages that do not reflect the distribution of properties over the whole remnant that SPI is able to realize. Even for a case which looks relatively `clean' such as \dem, SPI reveals substantial and irregular variation in all properties across the remnant. We show that this introduces biases for analyses which choose specific regions and assume they are spectrally homogeneous. Small regions 
cannot be assumed to be representative of the SNR as a whole, and larger regions 
average over the variations and miss small features. 

Relatively simple traditional approaches provide only very limited or no information about the wide range of physical conditions throughout the entire SNR. The most common approach of fitting the spectrum with a multi-component model will not reproduce the true variations (both spatial and spectral) of gas properties, unless an inordinately large number of components are used. Analyzing spectra of small regions covering the entire remnant is an alternative that is sometimes used \citep[e.g.][]{Yang2008,Lopez2013}; however, this requires extremely deep observations and still requires simple spectral models for each region. SPI does not require exceptionally deep observations, and allows for arbitrarily complex spectral models. It also removes any need to make assumptions about the spatial or spectral distribution of physical properties (e.g. spherical symmetry or number of different temperatures) while providing a way to measure more than simple `average' quantities. It can be used to make maps and distributions of any of the parameters present in the spectral models. 
SPI can therefore provide much more globally accurate conclusions about physical properties or the evolutionary state of a SNR, as compared to standard methods of X-ray analysis.

The nature of SPI makes it possible, and comparatively easy, to separate and study the properties of low EM gas, or to identify and isolate any subset of gas with given properties. Simple 1D simulations can enhance this capability by providing guidance on which parameters are most informative for a given SNR and by aiding the identification of different gas components, e.g. shocked ejecta, by considering the evolution of the SNR. In the case of \dem, we are able to separate emission from the ejecta based not on visual identification of surface brightness features, but rather on measured temperature and Fe abundances. This allows us to map and construct distributions of the ejecta temperature, ionization age, and other properties. We find, as expected from the 1D simulations, that this gas has lower ionization ages than the bulk of the SNR gas, as well as flatter O, Ne, and Fe distributions. For other SNRs, a similar procedure can be used to isolate and study components of interest. 

SPI also offers a unique method for identifying the CD without relying on surface brightness features. 2D simulations clearly predict a shell-like region of R-T instabilities at the CD and provide an estimate of its location and width. The largest variations in temperature are expected in this region. With SPI, it is possible to visualize such variations by mapping the standard deviation of the temperature distribution within each spatial bin. We find the largest temperature variations in a distinct hook-shaped feature that matches the expected size and location of the R-T unstable CD region. This demonstrates the power of SPI and its potential for extracting new physical insights from existing X-ray observations of SNRs.

Standard approaches are very valuable because the methods and uncertainties are generally very well understood, which is not yet the case for SPI. They are also very fast compared to SPI, which is resource-intensive and slow. However, SPI can provide unique and valuable insight, and is a good complement to traditional methods of X-ray analysis of SNRs. In future papers we will apply the SPI technique to several more SNRs for which observations with {\it XMM-Newton} exist in the database.

\ \\
\acknowledgments
This work was supported by NASA ADAP grant NNX15AH70G to Pennsylvania State University, with subcontracts to the University of Chicago and Northwestern University. Based on observations obtained with XMM-Newton, an ESA science mission with instruments and contributions directly funded by ESA Member States and NASA.

\vspace{5mm}
\facilities{XMM(EPIC)}

\bibliography{deml71_paper}

\end{document}